\documentclass[aps,prl,10pt,twocolumn,preprintnumbers,superscriptaddress,showpacs,floatfix,nofootinbib]{revtex4-2}
\usepackage{textcomp,gensymb}
\usepackage{hyperref}
\usepackage{graphicx}
\usepackage{amsfonts,amsmath,amssymb,bm,bbm}
\usepackage{color,soul}
\usepackage[dvipsnames]{xcolor}
\usepackage{slashed}
\usepackage[toc,page]{appendix}
\usepackage{flushend}
\usepackage{physics}
\usepackage{graphicx}
\usepackage{dcolumn}
\usepackage{float}
\usepackage[toc,page]{appendix}
\usepackage{accents}
\usepackage[compat=1.1.0]{tikz-feynman}

\hypersetup{
    pdfnewwindow=true,      
    colorlinks=true,       
    linkcolor=blue,          
    citecolor=blue,        
    filecolor=blue,      
    urlcolor=blue        
}

\begin{document}
\preprint{CETUP 2026-006}
\title{Resonant Trapping of Neutrinophilic Dark Matter in Supernovae}

\author{Christopher V. Cappiello}
\email{cappiello@wustl.edu}
\affiliation{Department of Physics and McDonnell Center for the Space Sciences, Washington University, St. Louis, MO 63130, USA}
\author{P. S. Bhupal Dev}
\email{bdev@wustl.edu}
\affiliation{Department of Physics and McDonnell Center for the Space Sciences, Washington University, St. Louis, MO 63130, USA}
\author{Amol V. Patwardhan}
\email{apatwardhan@reed.edu}
\affiliation{Department of Physics, Reed College, Portland, OR 97202, USA}

\begin{abstract}
 Supernova cooling provides some of the strongest constraints on light particles beyond the Standard Model, including sub-GeV dark matter (DM). While sufficiently large couplings to the Standard Model sector can trap these particles inside the supernova or cause them to decay before escaping, there typically exists an intermediate range of couplings in which their efficient production and subsequent escape lead to excessive energy loss, resulting in stringent constraints on new light particles. In this work, we demonstrate for the first time that, for certain types of DM-neutrino interactions, resonantly enhanced $s$-channel scattering with neutrinos can efficiently trap DM inside the supernova and can fundamentally alter the conventional supernova cooling constraints. In particular, we show that for mediator masses comparable to the supernova core temperature, this resonant scattering can prevent a substantial fraction of the produced DM from freely escaping, thereby significantly weakening, or even evading, the usual supernova cooling constraints on such models.
\end{abstract}
\maketitle

\section{Introduction}

It has long been recognized that anomalous cooling of supernovae can be used to study the existence and properties of particles beyond the Standard Model. In fact, the observation of SN1987A~\cite{Kamiokande-II:1987idp,Bionta:1987qt,Alekseev:1988gp} has been extensively used to set stringent constraints on axions and axion-like particles, new light mediators, and particle dark matter (DM)~\cite{Raffelt:1990yz, Raffelt:1996wa}. Supernova observations have also been used to set constraints on the properties of neutrinos, in particular non-standard neutrino interactions~\cite{Dicus:1982dk,Kolb:1987qy, Manohar:1987ec, Dicus:1988jh, Fuller:1988ega, Konoplich:1988mj, Kainulainen:1990bn, Shi:1993ee, Kolb:1996pa, Nunokawa:1997ct, Hidaka:2006sg, Hidaka:2007se, Fuller:2008erj, Blennow:2008er, Arguelles:2016uwb, 
Heurtier:2016otg, Das:2017iuj, Dighe:2017sur, Yang:2018yvk,Mastrototaro:2019vug, Shalgar:2019rqe, Suliga:2020vpz, 
Suliga:2020jfa, Akita:2022etk, Chang:2022aas, Chen:2022kal, Fiorillo:2022cdq, Ray:2023gtu,  Manzari:2023gkt,   Fiorillo:2023ytr, Chauhan:2023sci, Carenza:2023old, 
Bhattacharya:2023wzl,
Akita:2023iwq,
Lai:2024mse,Ray:2024jeu, Telalovic:2024cot, Suliga:2024nng, Ivanez-Ballesteros:2024nws,Chauhan:2025mnn}; see Ref.~\cite{Raffelt:2025wty} for a recent review. However, relatively little work has focused on the combination of these ideas: using supernovae to constrain interactions between DM and neutrinos~\cite{Manzari:2023gkt, Cappiello:2025tws, Chauhan:2025hoz, Tallur:2026fpq}.

In Ref.~\cite{Cappiello:2025tws}, we modeled supernova cooling due to DM production sourced by neutrino annihilation, assuming a fermion DM $\chi$ and a vector mediator $Z'$ with couplings of the type $\bar{\nu}\nu Z'$ and $\bar{\chi}\chi Z'$, which facilitate an $s$-channel production of DM via $\nu\bar{\nu}\to \chi\bar{\chi}$ and $t$-channel trapping via $\nu\chi\to \nu\chi$ scattering. 
  Due to the large neutrino density inside a supernova, and the high temperatures of these environments, we were able to derive stringent bounds on a wide range of parameter space, emphasizing in particular resonant production of DM through an on-shell MeV-scale mediator. However, these bounds do not extend to arbitrarily large/small couplings, because for larger couplings, the DM can be trapped by scattering with neutrinos, or can annihilate with itself back into neutrinos, while for very small couplings, the production process is no longer efficient. Although our analysis focused on vector mediators, we anticipate qualitatively similar behavior for a scalar mediator $\phi$ coupled to neutrinos and DM through interactions of the form \(\bar{\nu}\nu\phi\) and \(\bar{\chi}\chi\phi\), respectively.

\begin{figure}[t!]
    \centering
    \includegraphics[width=\linewidth]{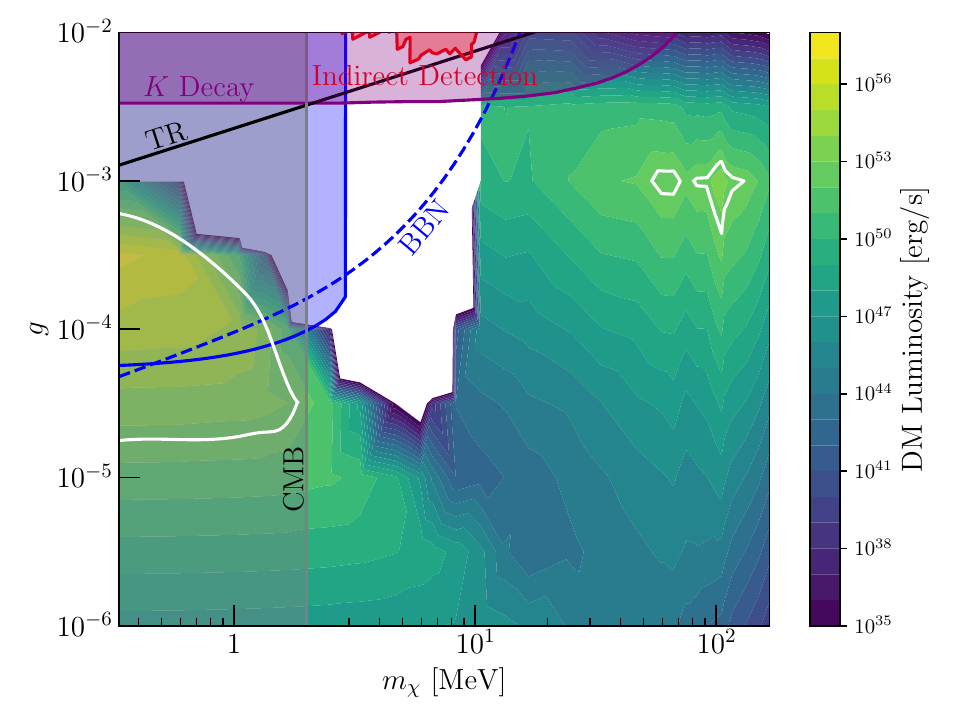}
    \caption{Total energy loss rate due to DM production (color scale) as a function of the DM mass $m_\chi$ and coupling $g$ [cf.~Eq.~\eqref{eq:Lag}], assuming a light mediator with $m_{\phi} = 3m_{\chi}$. The white contour delineates our SN1987A exclusion region, corresponding to an emissivity of \(5\times10^{52}~\mathrm{erg/s}\). For comparison, we display constraints from BBN (blue shaded and dashed regions)~\cite{Dev:2025tdv}, CMB (gray)~\cite{Li:2023puz}, DM indirect detection searches (red)~\cite{Arguelles:2019ouk}, and kaon decay (purple)~\cite{Dev:2024ygx}. The black curve (TR) indicates the parameter space consistent with the observed thermal DM relic abundance. See text for details.}
    \label{fig:limits}
\end{figure}

In this work, we study the supernova constraints for  interactions of the type $\bar{\chi}\nu\phi$ [cf.~Eq.~\eqref{eq:Lag}]. This leads to a $t$-channel production via neutrino annihilation, but $s$-channel scattering with neutrinos. 
 The most noticeable qualitative difference between this case and the previously studied case is that DM can no longer be produced resonantly, while it can resonantly scatter with neutrinos for light mediators. This results in trapping that is both very strong, and particularly dependent on the center-of-momentum energy of the DM-neutrino interaction. This leads to a fundamentally different supernova cooling constraint which has a highly nontrivial dependence on the mediator mass, and some parts of parameter space that are strongly constrained for the models featuring $t$-channel DM-neutrino scattering being completely unconstrained for the models featuring $s$-channel scattering; see Fig.~\ref{fig:limits}. 

\section{Dark Matter Production}

We consider a representative model of $\nu$-DM interactions in which a complex scalar mediator $\phi$ that carries a lepton number $L = 1$ mediates the following effective interaction:
\begin{equation}
    \mathcal{L}_\text{int} \supset - g \phi^* \bar\chi P_L \nu + \text{H.c.}
    \label{eq:Lag}
\end{equation}
Possible ultraviolet-completions of this type of interaction are discussed in Refs.~\cite{Farzan:2012ev,Arhrib:2015dez,Dutta:2020scq,Herms:2023cyy,Dev:2025tdv, Babu:2025czb}. For instance, such interactions naturally appear in popular radiative neutrino mass models, such as the scotogenic model~\cite{Tao:1996vb,Ma:2006km}. For simplicity, we assume a flavor-universal DM-neutrino coupling here. 
\begin{figure*}[htb!]
    \centering    \includegraphics[width=0.99\textwidth]{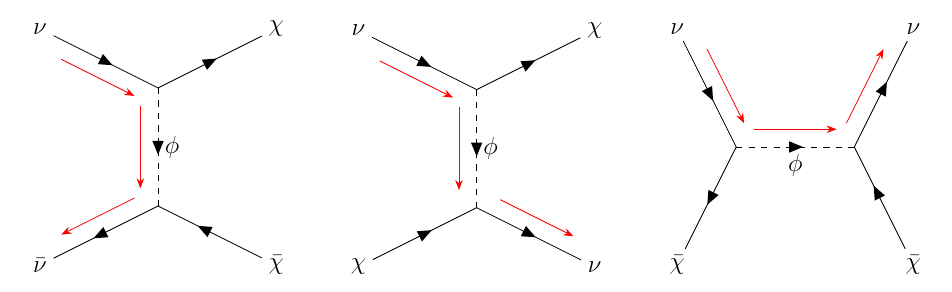} 
    \caption{Processes contributing to DM pair production (left) and trapping (center and right). The diagram on the right is the channel that exhibits resonant trapping. For the center and right diagrams, the charge-conjugated versions ($\bar\chi \bar\nu \to \bar\chi\bar\nu$, and $\bar\nu\chi \to \bar\nu\chi$) are also present, albeit not explicitly shown. In each diagram, the red arrows indicate the flow of lepton number. }
    \label{fig:feyn}
\end{figure*}
For the Lagrangian~\eqref{eq:Lag}, the DM production via neutrino annihilation and DM-neutrino scattering processes are shown in Fig.~\ref{fig:feyn}. The DM production cross section via $\nu_\alpha \bar \nu_\alpha \to \chi \bar \chi$ (for any neutrino flavor $\alpha$) is
\begin{align}  & \sigma_{\nu_\alpha\bar\nu_\alpha \to \chi \bar{\chi}} = 
    \frac{g^4}{16 \pi  s} \left[\sqrt{1 -\frac{4 m_\chi^2}{s}} \ \frac{2 (m_\chi^2-m_\phi^2)^2+m_\phi^2 s}{(m_\chi^2-m_\phi^2)^2+m_\phi^2 s}\right.\nonumber \\
    & \quad \left.+\frac{4 (m_\chi^2-m_\phi^2)}{s} \tanh ^{-1}\left(\frac{ \sqrt{s(s-4 m_\chi^2)}}{s-2 m_\chi^2+2 m_\phi^2}\right) \right] \, ,
    \label{eq:prod}
\end{align}
where for brevity we have set $m_\nu \to 0$. 

To compute the DM relic density and other cosmological and astrophysical bounds, we also need the cross section for DM annihilation to neutrinos, which is given by
\begin{align}
& \sigma_{\chi\bar{\chi} \rightarrow \nu_{\alpha}\bar{\nu}_{\alpha}}v = \frac{g^4}{16\pi s}\Bigg[\frac{ 2(m_{\chi}^2-m_{\phi}^2)^2+m_{\phi}^2s}{(m_{\chi}^2 - m_{\phi}^2)^2+m_{\phi}^2s}\nonumber \\
& + \frac{4(m_{\chi}^2 - m_{\phi}^2)}{s\sqrt{1 - \frac{4m_{\chi}^2}{s}}}\tanh ^{-1}\left(\frac{ \sqrt{s(s-4 m_\chi^2)}}{s-2 m_\chi^2+2 m_\phi^2}\right)\Bigg] \, .
\end{align}

For DM-$\nu$ scattering, there are two contributions: the differential scattering cross section for $\chi\nu \to \chi\nu$ is
\begin{align}
    \frac{{\rm d}\sigma_{\nu\chi\to \nu\chi}}{{\rm d}t} = \frac{g^4}{32 \pi (s-m_\chi^2)^2} \frac{(m_\chi^2-u)^2}{(m_\phi^2-u)^2} \, ,
    \label{eq:scat1}
\end{align}
whereas for $\chi\bar\nu \to \chi\bar\nu$, we get
\begin{align}    \frac{{\rm d}\sigma_{\bar\nu\chi\to \bar\nu\chi}}{{\rm d}t} = \frac{g^4}{32 \pi (s-m_\chi^2)^2} \frac{(s - m_\chi^2)^2}{(s - m_\phi^2)^2 + m_\phi^2 \Gamma_\phi^2} \, ,
\label{eq:scat2}
\end{align}
where $\Gamma_\phi$ is the total width of the mediator, given by 
\begin{align}
    \Gamma_{\phi} = 3\times\frac{g^2m_{\phi}}{16\pi}\left(1 - \frac{m_{\chi}^2}{m_{\phi}^2}\right)^2 \, ,
\end{align}
with the factor of 3 taking into account 3 neutrino flavors in the decay $\phi \to \nu\bar{\chi}$. We assume $m_\phi>m_\chi$ so that the decay $\phi \to \nu\bar{\chi}$ is kinematically allowed, and at the same time, the decay $\chi \to \nu\phi^*$ is forbidden which is essential for the stability of the DM. In Eqs.~\eqref{eq:prod}-\eqref{eq:scat2},  $s$, $t$, $u$ are the usual Mandelstam variables, and the cross section expressions are consistent with those given in  Ref.~\cite{Dev:2025tdv}. 

Note that Eq.~\eqref{eq:scat2} has an $s$-channel resonance, which is the key to resonant trapping. Moreover, since $\nu$ and $\bar\nu$ do not have identical distributions inside the protoneutron star (PNS), the trapping will be asymmetric across $\chi$ and $\bar\chi$, leading to asymmetric DM accumulation in the PNS over time. This can have interesting implications that we do not consider in the present work.

In principle, the process $\nu \bar{\nu} \rightarrow \phi \phi^*$ is also allowed, and could contribute to supernova cooling. However, in the parameter space we consider, the decay length of the mediator is very short compared to the radius of the neutrinosphere. For instance,
\begin{equation}
    \lambda_{\rm Decay} = (4.2~{\rm cm}) \times \beta \gamma \left(\frac{g}{10^{-5}}\right)^{-2} \left(\frac{m_{\phi}}{\rm MeV}\right)^{-1} \,,
\end{equation}
assuming $m_{\phi} = 3m_{\chi}$. Here $\beta\lesssim 1$ is the DM velocity and $\gamma$ is its Lorentz boost factor which roughly varies between 1 and 200 in the parameter space shown in Fig.~\ref{fig:limits}. Thus, we find that $\lambda_{\rm Decay}\ll R_\nu\simeq 40$ km, i.e., the mediator always decays deep within the neutrinosphere. 

Following Ref.~\cite{Cappiello:2025tws}, we compute the energy loss rate due to DM production from neutrino annihilation as
\begin{align}
    \dot{E}_0 = &\frac{8\pi^2}{(2\pi)^6}\int {\rm d}E_1{\rm d}E_2{\rm d}\cos{\theta}f_1f_2E_1^2E_2^2(E_1 + E_2) 
    \nonumber \\
    & \qquad \qquad \qquad \times \sigma_{\nu \bar{\nu} \rightarrow \chi \bar{\chi}}v_{\rm rel}\,.
    \label{eq:emis}
\end{align}
Here, $E_{1,2}$ are the energies of the annihilating neutrino and antineutrino, $v_{\rm rel}\simeq \sqrt{2(1-\cos\theta)}$ is the relative velocity between them, and $f_{1,2}$ are their Fermi-Dirac distributions, given by
\begin{equation}
    f_i = \frac{1}{1 + e^{(E_i \pm \mu(r))/T(r)}}\,,
\end{equation}
with $\mu(r)$ being the chemical potential as a function of the radial distance from the center of the supernova core, and the minus (plus) sign being used for neutrinos (antineutrinos). 

We characterize the neutrino medium using data from a one-dimensional core-collapse supernova simulation of a  PNS with a \(20~M_\odot\) progenitor~\cite{jankasim}. The simulation provides radial profiles of the density, temperature, and chemical potential at multiple post-bounce times. Consequently, through the radial dependence of the temperature and chemical potential, \(\dot{E}_0\) in Eq.~\eqref{eq:emis} acquires an implicit dependence on the radial coordinate, as we discuss explicitly below. The DM emission is found to be largest at a radius of approximately 8 km, reaching its maximum around 1 s after core bounce~\cite{Cappiello:2025tws}. We therefore adopt the simulation profiles at 1 s post-bounce for deriving our constraints, as this epoch corresponds to the most efficient DM production. Interestingly, the radius of peak emission also coincides with the location of the maximum temperature at this time. 

\section{Resonant Trapping}

After a DM particle is produced within the neutrinosphere, it must propagate out of the neutrinosphere, and in the process it may scatter with neutrinos or antineutrinos, or annihilate with other DM particles that have been produced by neutrino annihilation. Following the approach of Ref.~\cite{Cappiello:2025tws}, we only consider DM that escapes the neutrinosphere without scattering for the energy loss condition. We therefore write the volumetric emissivity of DM from a point within the neutrinosphere as
\begin{equation}
    \dot{E}(r) = \dot{E}_0(r)\exp\left(-\int_r^{\infty}\frac{\mathrm{d}r'}{\lambda(r')}\right)\,.
\end{equation}
Here, $\lambda$ is the mean free path of the DM (which accounts for both scattering with neutrinos and self-annihilation), computed via
\begin{equation}
    \lambda^{-1}(E_{\chi},\textbf{r}) = \frac{2\pi}{\beta}\int\sum_\alpha\Gamma_\alpha(E_{\chi},\theta,\phi)\mathrm{d}\cos\theta\, , 
\end{equation}
where $\Gamma_\alpha$ is the scattering rate for a DM particle with neutrinos of flavor $\alpha$. 

In practice, we neglect DM trapping due to annihilation, due to the fact that the produced DM density is small compared to the neutrino density in the neutrinosphere. Moreover, in the light mediator case, the scattering cross section with neutrinos has an $s$-channel resonance, while the annihilation cross section with DM does not. We follow Ref.~\cite{Cappiello:2025tws} in defining the scattering rate 
\begin{equation}
\Gamma_{\alpha}(E_\chi,\theta,\phi) = \frac{n_{\nu_{\alpha}} v_{\rm rel}(E_\chi,\theta)}{4\pi  [-2T^3{\rm Li}_3(-e^{\mu/T})]} \int \textrm{d}E_{\nu}E_{\nu}^2f(E_{\nu}) \,\sigma_{\textrm{PB}}\,,
\end{equation}
in terms of the Pauli-blocked cross section 
\begin{equation}
    \sigma_{\textrm{PB}} = \int [1-f_{\alpha}(E_{\rm out},\mu,T)]\frac{\textrm{d}\sigma}{\textrm{d}t} \textrm{d}t\,,
\end{equation}
and the relative velocity between the DM and neutrino  
\begin{equation}
    v_{\rm rel}(E_\chi,\theta) = \sqrt{1 - 2\beta \cos\theta + \beta^2}\,.
\end{equation}
The Pauli-blocked cross section accounts for the fact that neutrino degeneracy suppresses the probability of scattering an ambient neutrino to a final energy state that is already occupied.

\begin{figure}
    \centering
    \includegraphics[width=\linewidth]{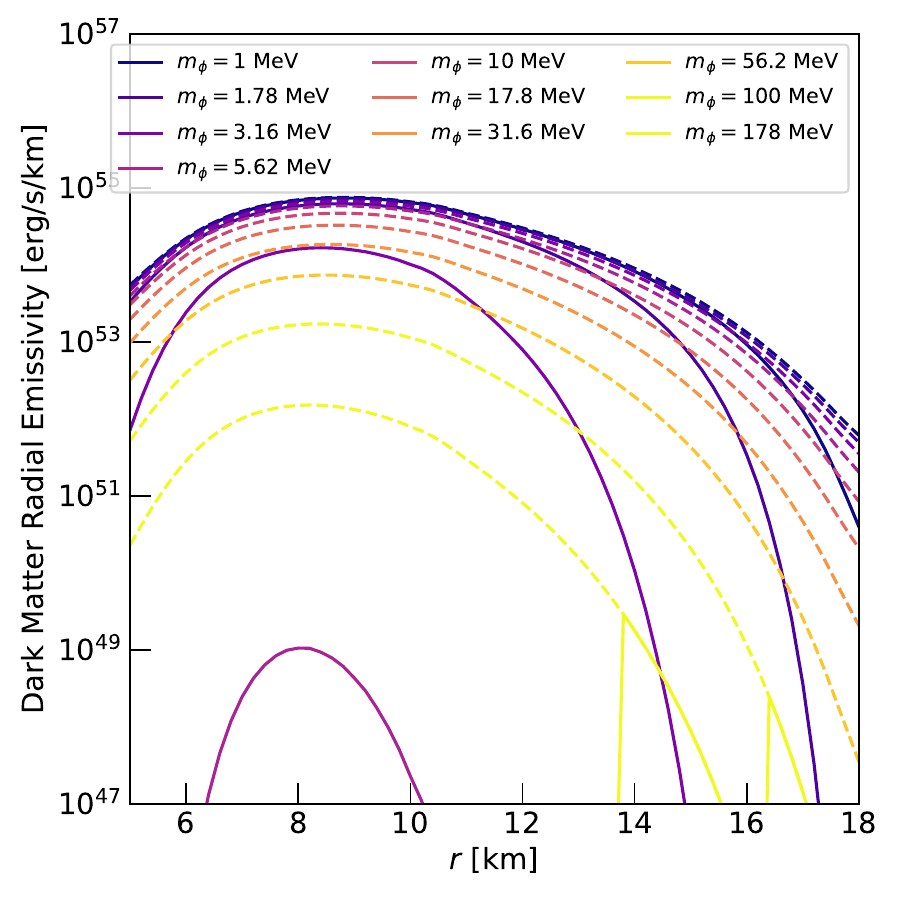}
    \caption{DM emissivity profile without trapping (dashed) and with trapping (solid) for different mediator masses, where we have set $m_{\chi} = m_{\phi}/3$ and $g = 10^{-4}$. This coupling and choice of masses illustrate the behavior of the emissivity across the exclusion region in Fig.~\ref{fig:limits}, and explains why there is no exclusion region at a few MeV.}
    \label{fig:EmissivityandTrapping}
\end{figure}

Fig.~\ref{fig:EmissivityandTrapping} shows the energy that escapes from the neutrinosphere due to DM production per second per radial km, for a mixed mass ratio $m_{\phi}/m_\chi = 3$ and coupling $g = 10^{-4}$. The dashed lines show the result when trapping is neglected, while the solid lines show the result when trapping is included. When the mediator mass is large, it is difficult to reach $s = m_{\phi}^2$ in DM-neutrino scattering, meaning that resonant scattering can only occur for DM produced near 8 km, where the temperature profile of the neutrinosphere peaks. As the mediator mass gets smaller, lower temperature is required to reach this resonance, meaning that DM produced across a wider range of radii will be trapped by resonant scattering. 

\section{Results}

\begin{figure}[t!]
    \centering
    \includegraphics[width=\linewidth]{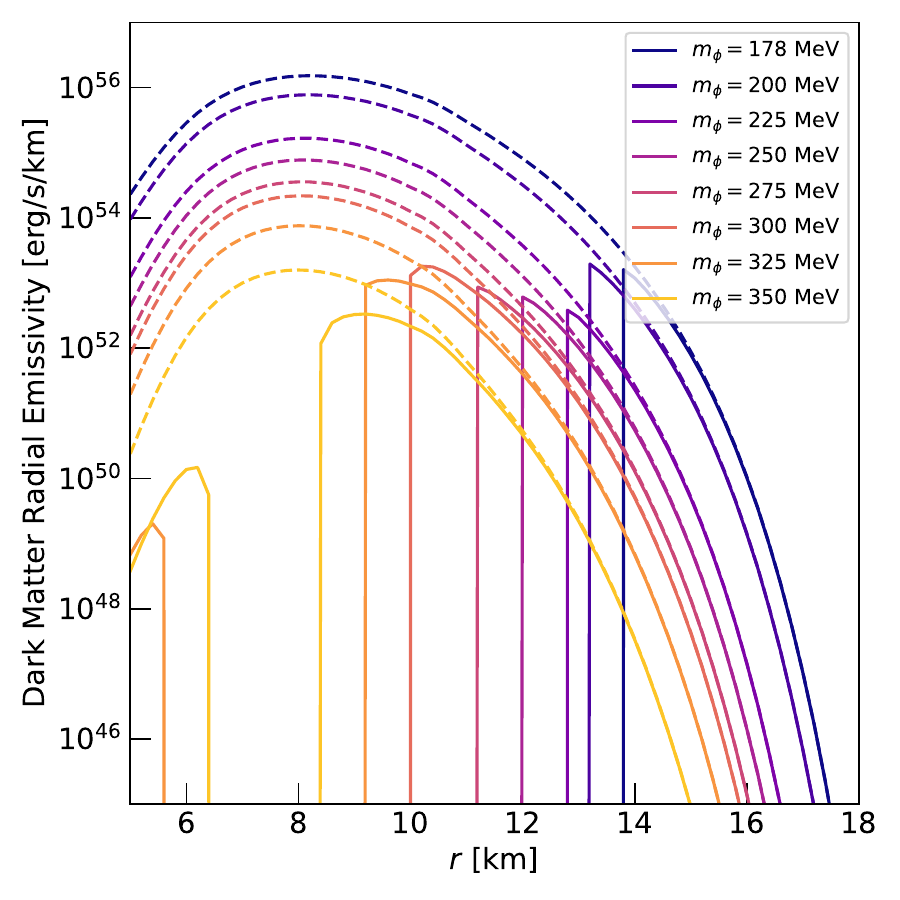}
    \caption{Same as Fig.~\ref{fig:EmissivityandTrapping}, but focused on mediator masses of a few hundred MeV, and with $g = 10^{-3}$. This choice of coupling is made in order to focus on the small excluded regions toward the right-hand side of Fig.~\ref{fig:limits}. }
    \label{fig:EmissivityandTrappingHighMass}
\end{figure}

In Fig.~\ref{fig:limits}, we show the DM luminosity as a function of the coupling and mediator mass $m_{\phi}$, assuming that $m_{\phi} = 3m_{\chi}$. In white, we denote the exclusion regions derived by requiring that the DM luminosity not exceed $5 \times 10^{52}$ erg/s ---  the so-called Raffelt criterion for SN1987A~\cite{Raffelt:1990yz}. We see that there is a large exclusion region at small $m_{\phi}$, and a somewhat small exclusion region for mediator masses of a few hundred MeV, but there is a large gap in the exclusion region between about 5 MeV and 100 MeV. The shape of this exclusion region can be understood as follows: when the mediator mass is larger than a few hundred MeV, the DM is heavier than roughly 100 MeV, making it difficult to produce efficiently, which sets the maximum DM mass that can be probed. For mediator masses between about 100 MeV and a few hundred MeV, the DM can be produced in large enough quantities to noticeably cool the supernova, but the mediator mass is still too large for resonant trapping to be an overwhelming effect; see Fig.~\ref{fig:EmissivityandTrapping} and a zoomed-in version in Fig.~\ref{fig:EmissivityandTrappingHighMass} focused on the high-mass regime. When the mediator mass is less than 100 MeV, resonant trapping becomes so efficient that the exclusion region completely disappears, leading to the gap in the exclusion region between about 5 MeV and 100 MeV. However, once the mediator mass is small enough, of order a few MeV, the typical neutrino and DM energies within the core of the supernova are so large that they can no longer resonantly scatter efficiently, leading to the exclusion region at small mass. Fig.~\ref{fig:EmissivityandTrappingHighMass} focuses on the range of mediator masses corresponding to the small exclusion regions at high mass in Fig.~\ref{fig:limits}. The emissivity is generally lower than in Fig.~\ref{fig:EmissivityandTrapping}, and more crucially, the peak emissivity displays non-monotonic behavior with respect to the mediator mass around $m_{\phi} = 200$ MeV, resulting in the two separate exclusion regions at high mass.

For comparison, we also show in Fig.~\ref{fig:limits} the line in parameter space corresponding to the thermal relic abundance (denoted as `TR' in black), as well as several existing constraints as shaded regions. Bounds arising from BBN and from kaon decay to leptons plus missing energy are taken from Ref.~\cite{Dev:2025tdv}. Indirect detection bounds are based on neutrino line search data from Super-Kamiokande, KamLAND, and Borexino~\cite{Arguelles:2019ouk}. While the CMB bound from $N_{\rm eff}$ data was originally derived for a $t$-channel scattering model~\cite{Li:2023puz}, we estimate that the result is similar for the $s$-channel case considered here. For a comprehensive discussion of these constraints and  various other constraints (not relevant to the parameter space shown here), see Ref.~\cite{Dev:2025tdv}. 

\begin{figure}[t!]
    \centering
    \includegraphics[width=\linewidth]{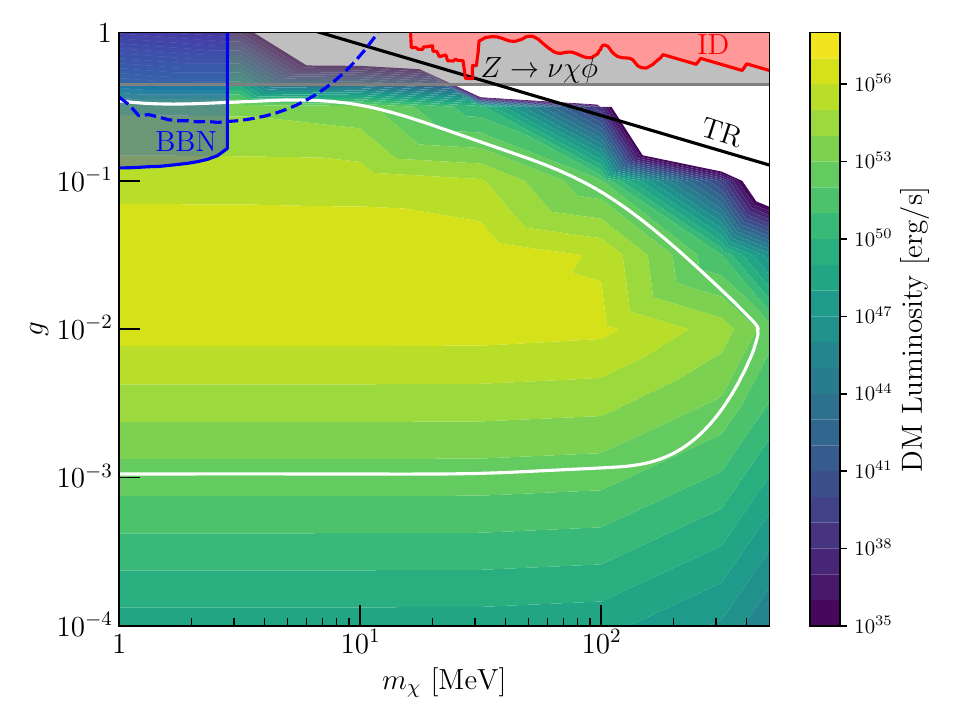}
    \caption{Same as Fig.~\ref{fig:limits} but for the heavy mediator case. Here $m_\phi=10$ GeV.}
    \label{fig:10GeVMediator}
\end{figure}

To see the dependence on the mediator mass, in Fig.~\ref{fig:10GeVMediator}, we present results for the heavy mediator case with a representative value of $m_{\phi}$ = 10 GeV, heavy enough that resonant effects may be neglected, and where in fact the theory maps directly onto the effective operator treatment presented in Ref.~\cite{Manzari:2023gkt}. In this case, we recover the well-known shape of the supernova exclusion region.

\section{Discussion and Conclusion}
  Although we have considered a scalar mediator for concreteness, we expect similar results for a vector mediator,  as long as the coupling is of the type $\bar{\chi}\nu Z'$ as in Eq.~\eqref{eq:Lag}. 
  
In this work, we have assumed that any DM that scatters after it is produced does not carry away any of its energy, i.e. it does not contribute to cooling. This is a somewhat conservative treatment, because although DM will typically lose energy when scattering, it could still escape and carry away some fraction of its initial energy. However, the cross section becomes so large in the resonant regime that we expect this to be a small effect: as can be seen in Figs.~\ref{fig:EmissivityandTrapping} and~\ref{fig:EmissivityandTrappingHighMass}, the probability of DM not scattering becomes extremely small once the resonant regime is reached, i.e. the mean free path is very short, so the probability of escaping after only a few collisions is also extremely small.

We also have not studied the effect of DM transferring energy between different regions of the supernova. This effect has been considered in the context of millicharged particle production~\cite{Fiorillo:2024upk}. It would be interesting to extend this treatment to DM-neutrino interactions.

Another interesting possibility is when the DM carries a lepton number, e.g. in certain models of sterile neutrino DM~\cite{Shi:1998km,Vogel:2025aut, Dev:2025sah}, instead of the mediator in Eq.~\eqref{eq:Lag}. This can lead to processes such as $\nu\bar{\chi}\to \bar{\nu}\chi$, thus transforming neutrinos into antineutrinos and vice-versa. Given that the $\nu_e$ number density can be a few orders of magnitude larger than the $\bar{\nu}_e$ number density in the central core region of the supernova~\cite{jankasim}, such processes could dramatically alter the observed supernova neutrino spectra and lead to additional constraints on DM-neutrino interactions. This will be studied in a follow-up work. 

In conclusion, we have shown that the supernova cooling constraints on DM-neutrino interactions depend on the type of interaction being considered. While the $\bar{\nu}\nu\phi$ and $\bar{\chi}\chi\phi$-type interactions lead to the typical cooling bounds expected for light particles, the $\bar{\chi}\nu\phi$-type of interaction leads to fundamentally different behavior, and significantly weaker or even vanishing constraints for ${\cal O}$(MeV) mediators. This provides a new mechanism for evading supernova constraints, which is qualitatively different from arguments based a different explosion mechanism itself~\cite{Bar:2019ifz} or DM self-trapping mechanism~\cite{Sung:2021swd}.    

\acknowledgments
\section*{Acknowledgments}
We are grateful to Hans-Thomas Janka for facilitating access to the numerical supernova simulation data maintained in the Garching Core Collapse Supernova Archive. The simulation data employed in this study originate from the work of Robert Bollig and Lorenz H\"udepohl using the \texttt{PROMETHEUS-VERTEX} code. B.D. thanks Kaladi Babu, Bhaskar Dutta, Doojin Kim, Shao-Ping Li, 
Deepak Sathyan, Kuver Sinha, Xunjie-Xu, and Yongchao
Zhang for useful discussions. A.V.P. thanks Sanjay Reddy for helpful discussions. C.V.C. was generously supported by Washington University in St.~Louis through the Edwin Thompson Jaynes Postdoctoral Fellowship. The work of B.D. was partly supported by the US
Department of Energy under grant No.~DE-SC0017987.  C.V.C. and B.D. 
wish to acknowledge the Center for Theoretical Underground Physics and Related Areas 
(CETUP*) and the Institute for Underground Science at SURF for hospitality and for
providing a stimulating environment, where part of this work was done.

\bibliography{main}

\end{document}